\documentclass[aps,prl,twocolumn,superscriptaddress]{revtex4}
\usepackage{amsmath,amssymb}
\usepackage{graphicx}
\usepackage{bm}
\usepackage{color}

\newcommand{\slr}{$(T_1T)^{-1}$}
\newcommand{\slrs}{$(T_{1s}T)^{-1}$}
\newcommand{\slrl}{$(T_{1\ell}T)^{-1}$}
\newcommand{\ybco}{YBa$_2$Cu$_3$O$_y$}
\newcommand{\ycbco}{Y$_{1-z}$Ca$_{z}$Ba$_2$Cu$_3$O$_y$}
\newcommand{\lsco}{La$_{2-x}$Sr$_x$CuO$_4$}

\begin{document}


\title{Evidence of a critical hole concentration in
underdoped \ybco\ single crystals revealed by $^{63}$Cu NMR}


\author{S.-H. Baek}
\email[]{sbaek.fu@gmail.com}
\affiliation{IFW-Dresden, Institute for Solid State Research,
PF 270116, 01171 Dresden, Germany}
\author{T. Loew}
\affiliation{Max-Plank-Institut f\"ur Festk\"orperforschung,
Heisenberg-strasse 1, 70569 Stuttgart, Germany}
\author{V. Hinkov}
\affiliation{Max-Plank-Institut f\"ur Festk\"orperforschung,
Heisenberg-strasse 1, 70569 Stuttgart, Germany}
\affiliation{Quantum Matter Institute, University of British Columbia,
2355 East Mall, Vancouver V6T 0A5, Canada}
\author{C. T. Lin}
\affiliation{Max-Plank-Institut f\"ur Festk\"orperforschung,
Heisenberg-strasse 1, 70569 Stuttgart, Germany}
\author{B. Keimer}
\affiliation{Max-Plank-Institut f\"ur Festk\"orperforschung,
Heisenberg-strasse 1, 70569 Stuttgart, Germany}
\author{B. B\"{u}chner}
\affiliation{IFW-Dresden, Institute for Solid State Research,
PF 270116, 01171 Dresden, Germany}
\affiliation{Institut f\"ur Festk\"orperphysik, Technische Universit\"at
Dresden, 01062 Dresden, Germany}
\author{H.-J. Grafe}
\affiliation{IFW-Dresden, Institute for Solid State Research,
PF 270116, 01171 Dresden, Germany}

\date{\today}

\begin{abstract}
We report a $^{63}$Cu NMR investigation in detwinned \ybco\ single crystals,
focusing on the highly underdoped regime ($y=6.35$---6.6).
Measurements of both the spectra and the spin-lattice relaxation
rates of $^{63}$Cu uncover the emergence of static order at a well-defined onset
temperature $T_0$ without a known order parameter as yet.  While $T_0$ is rapidly
suppressed with increasing hole doping concentration $p$, the spin pseudogap
was identified only near and above the doping content at which
$T_0\rightarrow 0$.
Our data indicate the presence of a critical hole
doping $p_c\sim 0.1$, which may control both the static
order at $p<p_c$ and the spin pseudogap at $p>p_c$.
\end{abstract}




\maketitle

The superconducting copper-oxides (cuprates) in the underdoped regime
feature unusual states of matter, such as a pseudogap (PG), density-wave
order (stripes), and the coexistence of magnetism and superconductivity.
Debates regarding the origin and the precise nature of those
phases or related phenomena like a reconstruction of the Fermi surface
at a quantum critical point are still
ongoing actively \cite{norman05, vojta09, taillefer09}.
In particular, interest in the underdoped \ybco\ (YBCO$_y$) has been
revived in recent years, during which significant progress in
understanding those subjects has been made through experimental observations
of an electric liquid crystal (ELC) or nematic phase \cite{ando02,hinkov08}, quantum
oscillations above a critical doping
\cite{doiron07, sebastian10, leboeuf11}, and field-induced charge
stripe order \cite{laliberte11,wu11}.

Motivated by the recent literature and by the available high-quality single
crystals of underdoped YBCO$_y$, we carried out a $^{63}$Cu NMR study of YBCO$_y$
to elucidate the underlying physics in
the highly underdoped region of the compound on a microscopic level.
While nuclear magnetic resonance (NMR) is a powerful local probe,
so far, the majority of the NMR studies on the planar Cu in YBCO$_y$
has been performed on nearly optimal or slightly
underdoped regions \cite{rigamonti98,walstedt08}, largely due to strong
magnetism which causes
complicated static and dynamic effects on NMR parameters.
In this Letter, we show that a critical
hole doping $p_c$ exists in the $p$-$T$ phase diagram of YBCO$_y$ beneath the
superconducting (SC) dome at $p\sim 0.1$,
below which a static order sets in and above which a spin pseudogap (PG) opens
up in the low-energy spin excitation spectrum.

The growth and characterization of detwinned YBCO$_y$ single
crystals are described in Refs. \cite{haug10,supple}. The single
crystals investigated here have $y=6.35$ ($T_c=10$ K, $p=0.062$),
$6.4$ ($T_c=21$ K, $p=0.075$), $6.45$ ($T_c=35$ K, $p=0.082$),
$6.5$ (sample 1 with short ortho II correlation length
\footnote{For intermediate oxygen concentrations, superstructures
can be formed in the CuO chains. For $x$=6.5, every second chain
is empty (ortho II phase), and the correlation length specifies
the range of ordering.}: $T_c=53$ K, $p=0.106$; sample 2 with long
correlation length ($\sim100$ \AA): $T_c=61$ K, $p=0.114$), and
$6.6$ ($T_c=61$ K, $p=0.135$), where $p$ is the hole concentration
per planar Cu determined from the $c$-axis lattice constant
\cite{liang06}. $^{63}$Cu NMR spectra were obtained by integrating
averaged spin-echo signals as the frequency was swept, and the
spin-lattice relaxation rates, $T_1^{-1}$, were measured by
monitoring the recovery of the nuclear magnetization after a
saturation pulse.

\begin{figure}
\centering
\includegraphics[width=\linewidth]{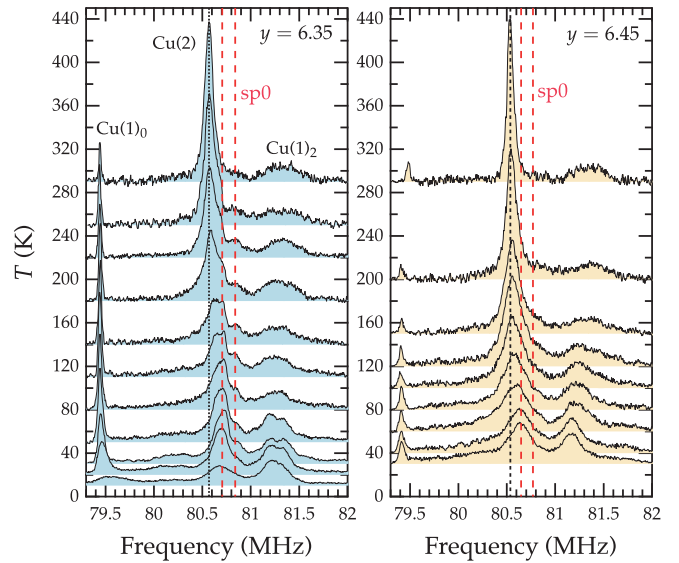}
\caption{\label{fig:spec} $^{63}$Cu NMR spectra in YBCO$_y$ at 7 T
along the $c$ axis for $y=6.35$ (left panel) and 6.45 (right
panel). Cu(1) denotes the Cu sites in the CuO chains, where the
subscripts 0 and 2 represent the number of neighboring oxygen ions
along the chain, and Cu(2) (dotted lines) denotes the planar Cu
sites \cite{takigawa91}. An emerging new Cu spectral feature sp0
consisting of two resonance lines indicated by two dashed lines is
clearly visible for $y=6.35$. The horizontal bar indicates the
region that was irradiated during a $T_1$ measurement. See
supplementary material for details and for other doping levels
\cite{supple}.}
\end{figure}

Fig.~1 shows $^{63}$Cu spectra for $y=6.35$ and 6.45 measured at
$H=7$ T applied along the $c$ axis. The leftmost sharp line and
the rightmost broad line in each panel are identified to arise
from the Cu(1)$_0$ and Cu(1)$_2$ sites, respectively, in the CuO
chains, where the subscript denotes the number of the nearest
neighboring oxygen ions \textit{along} the chain
\cite{takigawa91}. While the central line at room temperature
comes from the planar Cu(2) site \cite{takigawa91}, we find that
it evolves in a complicated way as $T$ is lowered, especially for
$y=6.35$ and $y=6.4$ \cite{supple}. The intensity of Cu(2) (dotted
line in Fig.~1) is strongly suppressed, and at the same time a new
feature sp0, consisting of two resonance lines indicated by two
dashed lines emerges at an onset temperature \cite{supple}.
Clearly, for $y=6.45$, sp0 occurs at a much lower temperature than
for $y=6.35$. The rapid suppression of Cu(2) with decreasing $T$
is attributed to the \textit{wipe-out effect}, as observed in
$^{63}$Cu NQR in \ycbco\ \cite{singer02},  which arises from a
slowing down of spin fluctuations \cite{hunt99, curro00}. In
contrast, the $T$-dependence of the $^{63}$Cu spectra for both
$y=6.5$ and 6.6 are almost identical without any signature of
either sp0 or wipeout of Cu(2), except for a moderate broadening
with decreasing $T$ \cite{supple}.
Interestingly, we observed the splitting of the Cu(1)$_2$ line at low $T$ for 
$y=6.35$, which may suggest that sp0 influences Cu(1)$_2$. 
In addition, it is noticeable that both sp0 and Cu(1)$_0$ broaden
significantly at low $T$ below 20 K for $y=6.35$, suggesting the occurrence of
glassy or incommensurate magnetic order. Note that such a broadening
could not be measured for $y=6.4$ and 6.45, because the magnetic order occurs deep in
the superconducting state where the NMR spectra significantly weaken and become complicated.

Fig.~2 shows the recovery of the $^{63}$Cu nuclear magnetization after a saturating pulse
as a function of delay time $t$ at various temperatures for $y=6.35$ and
6.45, measured at the Cu(2) line. Unexpectedly, we
found a step-like feature in the data, which indicates two $T_1$-processes.
The long $T_1$ process becomes discernible at 265 K for $y=6.35$ and at 120 K for
6.45, and its
fraction (i.e., $1-m(t)$ at the step)
increases rapidly with decreasing $T$.
Comparing with the $T$-dependence of the spectrum as shown in
Fig. 1, we find that the long $T_{1\ell}$ component arises when
the new spectrum sp0 becomes visible, suggesting that sp0 is due
to an emerging static phase featured by the long $T_{1\ell}$
\cite{supple}.
Since two $T_1$ processes are evident in the relaxation data, we
used a fitting function for magnetic relaxation of the central
transition for $I=3/2$ including two $T_1$ components,
\begin{equation}
\label{fit}
\begin{split}
m(t) = &\left[1-  a\left\{(1-w)\left(0.1e^{-t/T_{1s}}+
0.9e^{-6t/T_{1s}}\right)\right.\right. \\
&\;\;\;\;\;
+\left.\left.w\left(0.1e^{-t/T_{1\ell}}+0.9e^{-6t/T_{1\ell}}\right)\right\}\right], \\
\end{split}
\end{equation}
where $a$ is a fitting parameter which is ideally one for
saturation recovery. $T_{1s}$ ($T_{1\ell}$) are the short (long)
$T_1$ components, and $w$ is the fraction of the volume governed
by the long $T_{1\ell}$ process to the total volume. Solid curves
in Fig. 2 are fits to Eq. (\ref{fit}). At low $T$ ($\leq70$ K), it was
necessary to impose the stretching exponent $\beta$ in Eq. (1) for
the $T_{1\ell}$ recovery, which is indicative of a crossover to a
glassy magnetic phase.

\begin{figure}
\centering
\includegraphics[width=0.9\linewidth]{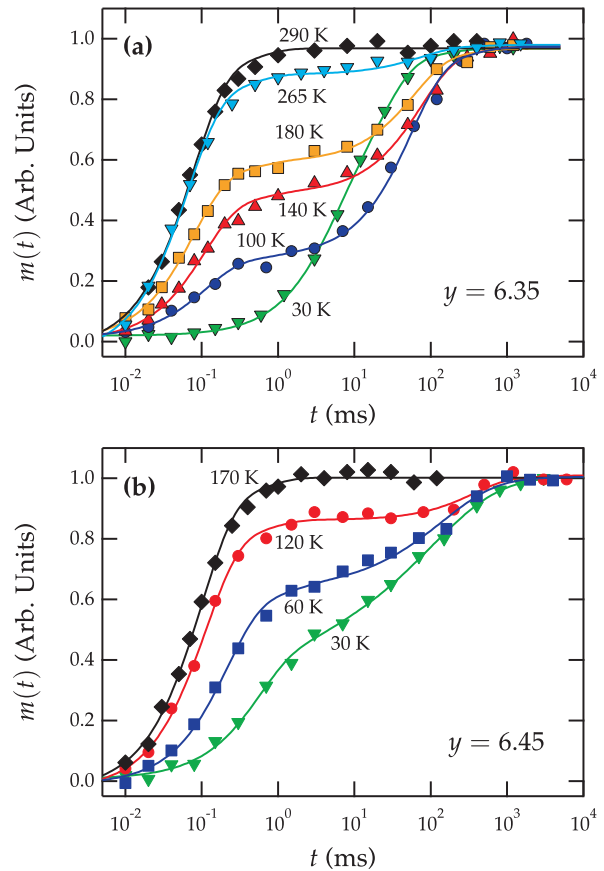}
\caption{\label{fig:rec} Recovery of the (normalized) nuclear magnetization $m(t)$
as a function of time $t$ for $y=6.35$ (a) and $y=6.45$ (b).
In (a), $m(t)$ deviates progressively from a
single $T_1$-process with decreasing $T$, revealing the long
$T_{1\ell}$-process which overwhelms the short one at low $T$.
Note that the short $T_{1s}$-component
is no longer detectable at 30 K.
In (b),
the $T_{1\ell}$-process is strongly suppressed i.e.,
it appears below $\sim 150$ K and a large portion of the $T_{1s}$-process
remains down to 30 K. Solid lines are fits to Eq.~(\ref{fit}). }
\end{figure}

The resulting \slrs, \slrl, and $w$ as a function of $T$ and
for all $y$ are presented in Fig.~3.
The results in YBCO$_{6.6}$ are compatible with data known so far
\cite{takigawa91}, revealing the spin pseudogap with the onset
$T^*\sim 145$ K. The spin-lattice relaxation rate probes the
gap solely in the spin excitation spectrum since
$(T_1T)^{-1}\propto \sum_\mathbf{q} A^2(\mathbf{q})
\chi''(\mathbf{q},\omega_0)$ where $A(\mathbf{q})$ is the hyperfine
coupling constant and $\omega_0$ the Larmor frequency. Therefore,
the onset temperature of the PG and its doping dependence obtained
by \slr\ can be significantly different from those obtained by the
Knight shift, which probes the spin response at $\mathbf{q}=0$ only
\cite{singer02}, and other techniques such as
ARPES and optical conductivity which probe the charge gap
\cite{norman05}. For YBCO$_{6.5}$, $T^*$ is reduced to $\sim$~110
K as indicated by arrows. We performed the measurement in
another YBCO$_{6.5}$ crystal which has a longer ortho-II correlation
length $\sim100$~\AA. For this sample, both $p$ and $T_c$ turns out to be
larger than the
sample with shorter correlation length, being consistent with slightly larger
$T^*$.
Upon further lowering $y$ to 6.45, \slrl\ increases with
decreasing $T$ reaching a plateau below $\sim150$ K, which is in
contrast to the pseudogap behavior reported in a similarly doped
compound \cite{berthier97}. \slrs\ of YBCO$_{6.35}$ and
YBCO$_{6.4}$ is further enhanced with no signature of the PG as
well, although $T_{1s}$ is not measurable below 100 K and 50 K,
respectively, due to limited experimental resolution [see Fig.
2(a)]. Note that the $T$-dependence of the data undergoes a
dramatic change as $y$ is reduced from 6.5 to 6.45, implying the
existence of a critical doping just below $y=6.5$.
\begin{figure}
\centering
\includegraphics[width=0.8\linewidth]{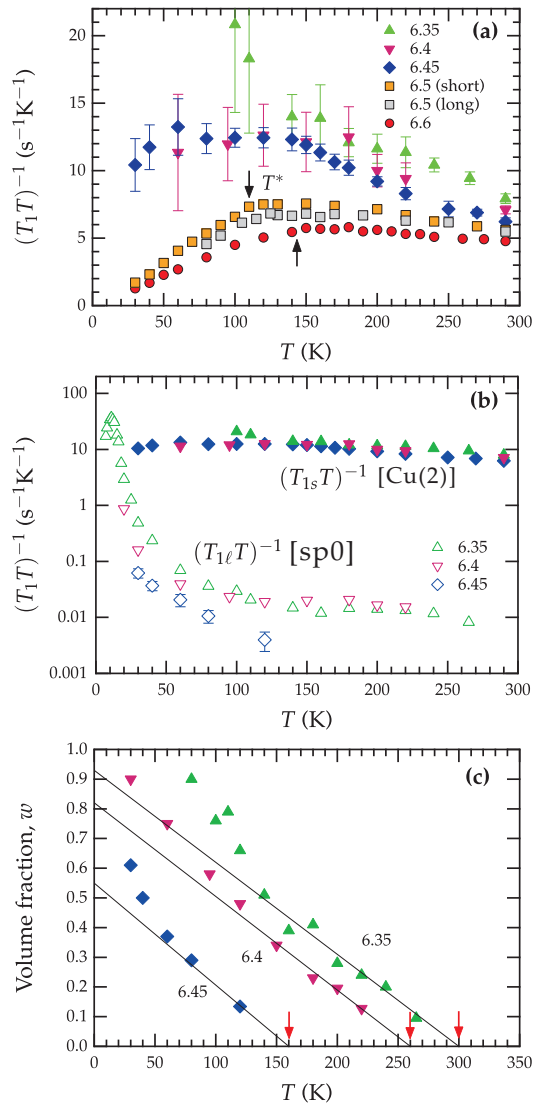}
\caption{\label{fig:T1} (a) Nuclear spin-lattice relaxation rates
divided by $T$, $(T_1T)^{-1}$ versus $T$ for the short $T_{1s}$.
The opening of the PG at
$T^*$ (denoted by arrows) was detected only for $y\ge6.5$, whereas
there are no signatures of the PG for $y\le 6.45$.
In (b), \slrl\ data are presented, together with \slrs\
for comparison. In YBCO$_{6.35}$, \slrl\ forms a local maximum at
$T_g\sim 11$ K, indicating a SG transition. This behavior is
significantly suppressed for $y=6.45$. (c) $T$-dependence of the
volume fraction $w$. The onset temperature $T_0$ is defined from
the values at which $w\rightarrow 0$ (down arrows).
}
\end{figure}

Fig. 3(b) shows the long relaxation rate \slrl\ versus $T$ which
arises from the sp0 line. For comparison, corresponding \slrs\
data of Cu(2) are also shown. ($T_{1\ell}$ is not detected for
$y=6.5$ and 6.6.) In YBCO$_{6.35}$, \slrl\ is almost constant at
high $T$, but it starts to rise steeply below $\sim 100$ K,
forming a sharp peak centered at $\sim11$K. These behaviors,
together with the significant broadening of the spectra at
low $T$,
lead to the conclusion that a spin freezing or spin-glass (SG)
transition occurs at a characteristic temperature $T_g \sim 11$ K
for $y=6.35$. We find that our data resembles the results of
$^{89}$Y NMR in \ycbco\ \cite{singer02} and $^{139}$La NQR/NMR in
\lsco\ \cite{chou93,julien99a}. Similar $T$-dependencies of
\slrl\ were also observed for $y=6.4$ and $6.45$.  Although we were
not able to identify the local maximum for these doping levels due
to
the higher $T_c$ which complicates the identification of $T_g$ by
NMR, one can obtain $T_g\sim5$ K for $y=6.45$ from the $\mu$SR
study of Y$_{1-x}$Ca$_{x}$Ba$_2$Cu$_3$O$_6$ which shows a very
similar $p$-dependence of $T_g$ \cite{niedermayer98}.

Fig. 3(c) shows the volume fraction $w$ as a function of $T$,
obtained from the fit of relaxation data $m(t)$ using Eq. (1). It should be
emphasized that the values of $w$ themselves have no quantitative
meaning, since they may depend on the frequency at which the
relaxation rates were measured. Moreover, the wipeout of
Cu(2) should lead to a significant increase of $w$, particularly
at low $T$, which is indeed thought to account for the rapid
increase of $w$ at low temperatures [see Fig. 3(c)]. Nonetheless,
the temperature at which $w\rightarrow 0$, i.e., where the $T_{1\ell}$
process vanishes with increasing $T$ should be unaffected by those
facts. Thus, with reasonable accuracy, one can define the onset
temperature $T_0$ from the values extrapolated to $w=0$, denoted
by arrows.

\begin{figure}
\centering
\includegraphics[width=\linewidth]{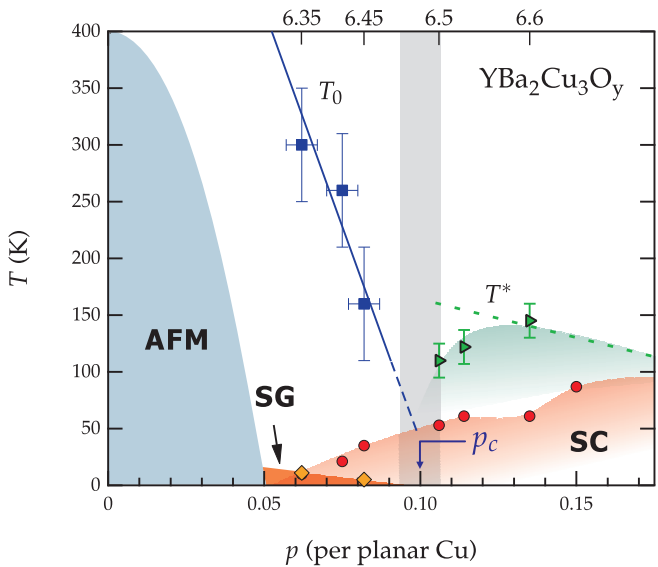}
\caption{\label{fig:phase}
Phase diagram of underdoped YBCO$_y$ in terms of hole concentration $p$.
$y$ values are shown on the top axis for convenience. AFM transition
is from Ref.~\cite{haug10}
and dotted line drawn for $T^*$ is estimated from previously known NMR results
\cite{rigamonti98,singer02}.  }
\end{figure}

From these results, we draw the $p$-$T$ phase diagram in Fig. 4.
The most striking feature is that $T_0$ falls rapidly
to zero at a hole concentration of $p\sim 0.1$ beneath the SC dome. At the same time,
the SG transition
temperature $T_g$ shows also similar doping
dependence, being terminated at $p$ where $T_0\rightarrow 0$.
These behaviors suggest
a close relationship between $T_0$ and $T_g$, collapsing to the same critical doping
$p_c\sim 0.1$. Interestingly, it turns out that $p_c$
is very
near the doping level at which the metal-insulator crossover (MIC) takes place \cite{sun04}
or the Fermi surface is reconstructed by density-wave order
\cite{leboeuf11,laliberte11,ghiringhelli12}.
The phase diagram is also in qualitative agreement with that
suggested by inelastic neutron scattering (INS) and muon spin rotation ($\mu$SR)
studies \cite{haug10,coneri10}.

A remarkable finding in our study is that the \textit{spin}
pseudogap is observed only near and above $p_c$ (i.e., $y\ge 6.5$)
[see Figs. 3(a) and 4]. Note that $T^*$ of YBCO$_{6.5}$ positioned
near $p_c$ is much lower than the value expected based on
extrapolation of the doping dependence established at higher $p$,
which is indicated by the dotted line.  Indeed, the sudden drop of
$T^*$ near $p_c$ in YBCO$_{6.5}$ appears as a crossover to the
absence of $T^*$ in YBCO$_{6.45}$, corroborating the INS results
in which the suppression of the low energy spin excitations (i.e.,
spin pseudogap) was not detected down to 5 K in YBCO$_{6.45}$
\cite{rossat-mignod91,hinkov08,li08b}. Such an abrupt suppression
of the spin pseudogap \textit{below a critical doping level} was
reported in NMR studies not only of similarly hole-doped \ycbco\
\cite{singer02} but also of the multi-layer cuprate
Ba$_2$Ca$_2$Cu$_3$O$_6$(F$_{y}$O$_{1-y}$)$_2$ \cite{shimizu11}. By
the same token, the fact that $p_c$ is in the vicinity of the MIC
\cite{sun04} agrees well with the disappearance of the quantum
oscillations below $\sim p_c$, as well as with the NMR result in
Bi$_2$Sr$_{2-x}$La$_{x}$CuO$_{6+\delta}$ that the ground state of
the PG is metallic \cite{kawasaki10a}. Therefore, we argue that
the spin PG phenomenon may be quantum critical, stemming from the
suppression of magnetism. In fact, such a strong competition
between magnetism and the PG gives a good account of the
observation that the PG is abruptly suppressed by substituting the
Cu(2) sites with $\sim 1$\% Zn impurities around which a local
moment is induced on the Cu(2) sites in YBCO$_{6.7}$
\cite{julien00} and YBa$_2$Cu$_4$O$_8$ \cite{zheng93}.

While the nature of static order which appears at $T_0$ is
unclear yet, one may consider $T_0$ as the onset of stripe-like charge
modulation in the plane, possibly induced by the end Cu ion of the
oxygen-filled chains, Cu(1)$_1$.
In fact, as effective impurities, Cu(1)$_1$ sites can cause
Friedel-like oscillations \cite{yamani06}, which may in turn induce the
stripe-like charge modulation in the plane.
This would be consistent with a higher $T_0$
at lower $p$, where the number of Cu(1)$_1$ is higher.
The charge modulation scenario agrees well with signatures of broken
rotational symmetry observed in
resistivity \cite{ando02} and INS \cite{hinkov08, haug10} measurements at
$p<p_c$. In this case, the planar
Cu(2) sites should be differentiated into two spatially modulated distinct regions.
In terms of stripes, the short $T_{1s}$-yielding region is naturally related to spin stripes
which consist of the localized Cu(2) spins, while charge stripes may yield the
long $T_{1\ell}$ if the spin contributions to the relaxation rates were almost
quenched.
Although the stripe-like charge modulation accounts for our
results to large extent, it remains a question whether the two alternating
regions in the stripe structure could, in practice, result in the \slrs\ and \slrl\ values
that strikingly differ up to three orders of magnitude.

In summary, we performed a systematic $^{63}$Cu NMR study as a
function of doping and temperature in highly underdoped \ybco,
showing that static order, probably stripe-like, emerges at the onset
temperature $T_0$, being followed by glassy magnetic order at $T_g$.
The resulting phase diagram includes a critical hole doping
$p_c\sim 0.1$, at which both $T_0$ and $T_g$ fall to
zero. Another important finding is that the spin pseudogap was detected only
above $p_c$, suggesting that the spin pseudogap phase competes with the magnetic order
and/or the static order detected at $T_g$ and $T_0$, respectively.

We thank M. Vojta, M.-H. Julien, and N. J. Curro for
useful suggestions and discussion, and D. Paar for experimental help.
This work has been supported by the Deutsche Forschungsgemeinschaft (DFG)
through FOR 538 (Grant No. BU887/4) and SPP1458 (Grant No. GR3330/2).

\bibliography{mybib}

\end{document}